\documentclass[aps,preprint]{revtex4}
\usepackage{graphicx}

\def\beq{\begin{eqnarray}}
\def\eeq{\end{eqnarray}}
\def\non{\nonumber}
\def\la{\langle}
\def\ra{\rangle}

\begin{document}

\title{ Weak decays of $J/\psi$: the non-leptonic case}

\author{Yu-Ming Wang$^{1}$} \author{Hao Zou$^{1}$} \author{Zheng-Tao Wei$^{2}$}
\author{Xue-Qian Li$^{2}$} \author{Cai-Dian L\"{u} $^{1}$}

\affiliation{$^{1}$Institute of High Energy Physics, P.O. Box
 918(4), Beijing 100049, China}

\affiliation{$^{2}$Department of Physics, Nankai University, Tianjin
 300071, China}

 \date{\today}

\begin{abstract}

In our previous study, we calculated the transition from factors of
$J/\psi\to D^{(*)}_{(s)}$ using the QCD sum rules.  Based on the
factorization approximation, the obtained form factors can be
applied to evaluate the weak non-leptonic decay rates of $J/\psi\to
D^{(*)}_{(s)}+M$, where $M$ stands for a light pseudoscalar or
vector meson. We predict that the branching ratio for inclusive
non-leptonic two-body weak decays of $J/\psi$ which are realized via
the spectator mechanism, can be as large as $1.3 \times 10^{-8}$, in
particular, the branching ratio of $J/\psi\to D^{*\pm}_s+\rho^\mp$
can reach $5.3 \times 10^{-9}$. Such values will be marginally
accessed by the ability of BESIII which will begin running very
soon.

\end{abstract}


\maketitle


\section{Introduction}

The decays of $J/\psi$ are dominated by strong and electromagnetic
interactions via $c\bar c$ annihilating into intermediate gluons and
photon at s-channel. By contrast, the weak decays, due to smallness
of the strength of weak interaction, are rare processes. Under the
spectator approximation, one of the charm quark or anti-charm quark
in $J/\psi$ decays into light quarks, and the decay rate of a charm
quark (anti-quark) is proportional to $G_F^2m_c^5$ where $G_F$ is
the Fermi coupling constant. Numerically the total branching ratio
of weak decays was estimated to at the order of $10^{-8}$
\cite{Sanchis-Lonzano}. Recently, due to remarkable improvements of
experimental instruments and techniques people turn their interests
onto these rare processes from both experiment
\cite{BES-semi,BES-non} and theory \cite{Verma 1,Verma 2,WZWLL}
sides. The forthcoming upgraded BESIII will be able to accumulate
more than $10^{10}$ $J/\psi$ per year \cite{BESIII}, which makes it
possible to marginally measure such weak decays in near future. More
important, such rare processes are  also particularly interesting
from the viewpoint of theory. On the one hand, it may provide
further accurate examination of the mechanism which is responsible
for the hadronic transition and fully governed by non-perturbative
QCD effects. One can also expect that such decays may offer a unique
opportunity to probe new physics beyond the standard model
\cite{zhangxm 1,zhangxm 2}, including the minimal supersymmetric
standard model, the extra dimension model, the two-Higgs doublet,
topcolor-assisted technicolor model etc , in the weak decay of
vector mesons. The reason is that in such rare decays, weak coupling
is rather weak and  new physics may have a chance to show up.

In a previous study, we presented a detailed analysis of the
semi-leptonic decays of $J/\psi$ \cite{WZWLL}, where the branching
ratios for such channels were estimated to be at order of $10^{-10}$
and hence is almost impossible to be observed at BESIII.

The fundamental ingredients involved in the semi-leptonic processes
are the transition form factors of $J/\psi\to D^{(*)}_{(s)}$, which
are evaluated in terms of the three-point QCD sum rules (QCDSR)
\cite{SVZ 1, SVZ 2, SVZ 3} in that work. Obviously, even though
while deriving the form factors our goal was to estimate the
branching ratios of semi-leptonic decays, under the factorization
approximation, they can be applied to study the non-leptonic decays.
Thus, we will take a step forward to investigate the exclusive
non-leptonic decays with focusing on two-body  processes.

\begin{figure}[tb]
\begin{center}
\begin{tabular}{ccc}
\includegraphics[scale=0.9]{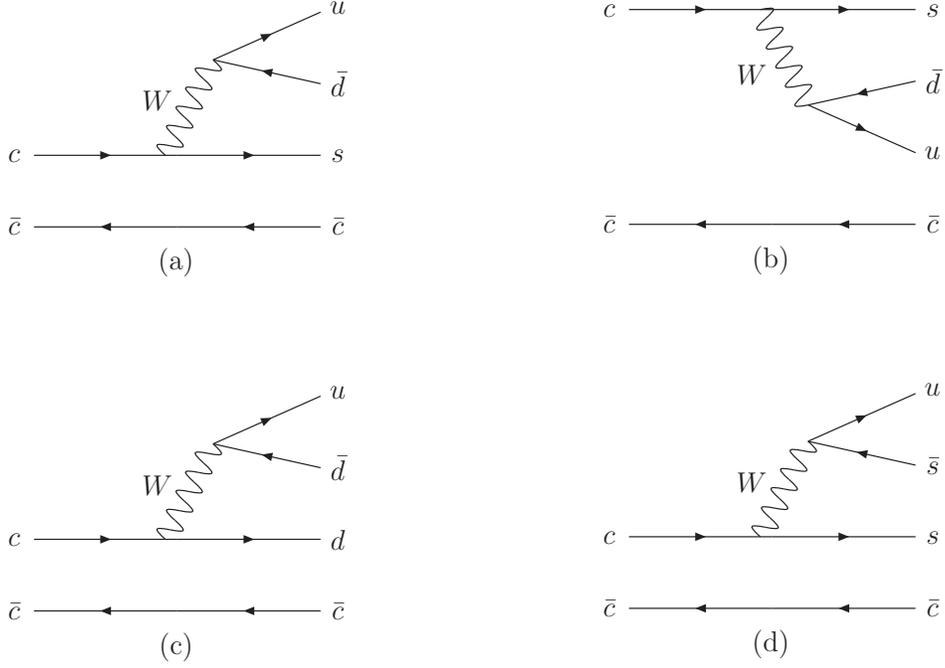}
\end{tabular}
\caption{Quark diagrams for non-leptonic weak decays of $J/\psi$.
(a) represents the color allowed processes; (b) represents the color
suppressed processes; (c), (d) represent single-Cabibbo suppressed
processes.} \label{f1}
\end{center}
\end{figure}

In this study, we will explore the non-leptonic decays $J/\psi\to
D^{(*)}_{(s)}+M$, where the final states contain a single charmed
meson and a light meson $M$, such as $\pi$, $K$, $\rho$, $K^*$ etc..
These weak decays are realized via the spectator mechanism that one
of the charm quark (anti-charm quark) acts as a spectator. In the
Standard Model, at the quark level, the Feynman diagrams for charm
quark decay are depicted in Fig. \ref{f1}. The anti-charm quark
decay can be obtained analogously by exchanging $c\leftrightarrow
\bar c$. The effective theory for hadronic weak decays have been
well formulated \cite{Buras}. The most difficult work is to
calculate the hadronic matrix elements which are governed by the
non-perturbative QCD dynamics.

The non-relativistic QCD   can simplify the picture by
phenomenologically handling some non-perturbative QCD effects and
has been widely applied to study some decay modes where heavy
quarkonium are involved. However, it does not help much for the
heavy-light mesons where relativistic effects may be significant.

The first order approximation for the derivation is the
factorization hypothesis, where the hadronic matrix element is
factorized into a product of two matrix elements of single currents
\cite{Factorization 1,Factorization 2,Factorization 3,Factorization
4,Factorization 5,Factorization 6}. In this scheme, one element can
be written in terms of the decay constant of the concerned meson
while the other is expressed by a few form factors according to the
Lorentz structure of the current and meson (a pseudoscalar or
vector, for example). The non-factorizable effects are incorporated
into the effective coefficients which are usually assumed to be
universal and determined by experiment (Only in some cases, they are
pertubartively calculable. In reality, these coefficients depend on
the concrete processes and differ case by case, but the variation
may be not very drastic.). For the weak decays of heavy mesons, such
factorization approach is verified to work very well for  the
color-allowed sub-processes. It is reasonable to believe that this
conjecture would be valid for $J/\psi$, at least for the processes
where the color-allowed sub-processes dominate. Thus, the study on
two-body non-leptonic decays offer an ideal ground to testify the
factorization hypothesis in the heavy quarkonium system and this
test may be more appealing than in decays of $D^{(*)}$ because
$J/\psi$ contains two heavy constituents. Moreover, they are of
great importance to discriminate various theoretical tools for the
evaluations of transition form factors.

The structure of this paper is as follows. In section II, the
factorization approach for the non-leptonic decays is introduced and
the formulations are given. In section III, after displaying the
inputs involved in this work explicitly, the numbers of branching
fractions for various $J/\psi\to D^{(*)}_{(s)}+M$ modes are
presented and comparisons of our numerical results with that
estimated in other theoretical models are also investigated at
length in this section. The final section is devoted to the
discussions and conclusions. It is noted that since most of the form
factors applied in this work were obtained in our previous work, we
generally refer the readers to it for some details of the derivation
and how to achieve the numerical values.

\section{Non-leptonic decays $J/\psi \to D_{(s)} +M$ in factorization approach}

For the non-leptonic weak decays of $J/\psi \to D_{(s)}+M$, the
standard method is integrating out the heavy $W-$boson and
obtaining a low energy effective Hamiltonian for $c$ quark decay
which is given by
\begin{eqnarray}
 \mathcal{H}_{eff}(c \to q u\bar q')={G_{F}\over\sqrt{2}}V^{*}_{cq}V_{uq'}\left(
 C_1Q_1+C_2Q_2\right),
\end{eqnarray}
where $q(q')$ represents the down type quarks $s$ and $d$;
$V^{*}_{cq}(V_{uq'})$ are CKM matrix elements; and the operators
$Q_1,~Q_2$ are respectively
\begin{eqnarray}
 Q_1=\bar q_{\alpha}\gamma_{\mu}(1-\gamma_5)c_{\alpha}~
  \bar q'_{\beta}\gamma^{\mu}(1-\gamma_5)u_{\beta}, \qquad
 Q_2=\bar q_{\alpha}\gamma_{\mu}(1-\gamma_5)c_{\beta}~
  \bar q'_{\beta}\gamma^{\mu}(1-\gamma_5)u_{\alpha}.
\end{eqnarray}
It should be pointed out that the penguin operators are neglected in
this work due to the smallness of  Wilson coefficients for such
operators, which also indicates that CP symmetry is well respected
within the accepted assumption.

With the free quark decay amplitude, we can proceed to calculate the
transition amplitudes for $J/\psi \to D+M$ at hadron level, which
can be obtained by sandwiching the free-quark operators between the
initial and final mesonic states. Consequently, the hadronic matrix
elements $\la DM|Q_i|J/\psi\ra$ which depend on the strong
interactions need to be computed. The evaluation is indeed the main
challenge in the heavy flavor physics due to our poor knowledge with
respect to the non-perturbative QCD. Owing to the painstaking
efforts in theory, several systemic approaches for hadronic $B$
decays have been explored based on the expansion in small parameters
\cite{Beneke}. However, a systematic theoretical method concerning
 the open-charm decays  is still not available yet due to the fact
that the accessible charm quark mass is not so heavy  in reality. As
the first order approximation, we may be able to apply the vacuum
saturation approximation to factorize the four-quark operator matrix
elements $\la DM|Q_i|J/\psi\ra$. The consistency of the theoretical
prediction with data (may be available in the future) will serve as
an examination of such approximation in the heavy-quarkonium system
as mentioned in the introduction. To be more specific, the
factorization ansatz \cite{Factorization 1} states that the matrix
elements can be factorized into a product of two single matrix
elements of currents $\la M|J_1|0\ra\la D|J_2|J/\psi\ra$ where one
is parameterized by the decay constant of the emitted light meson
and the other is represented by the form factors responsible for the
transition  of $J/\psi$ into the recoiled charmed meson.

The decay constants for pseudoscalar ($P$) and vector ($V$) mesons
are defined as follows
 \beq \label{dc}
 \la P(q)|A_{\mu}|0\ra&=&-if_P q_{\mu}, \non\\
 \la V(q,\epsilon)|V_{\mu}|0\ra&=&f_Vm_V\epsilon^*_{\mu},
 \eeq
where the axial vector current $A_{\mu}$ represents $\bar
q_1\gamma_{\mu}\gamma_5q_2$ and the vector current $V_{\mu}$
represents $\bar q_1\gamma_{\mu}q_2$; $\epsilon$ is the polarization
vector of $V$. The matrix elements $\langle
D|\bar{q}\gamma_{\mu}(1-\gamma_5)c|J/\psi\rangle$ are parameterized
in terms of various form factors as \cite{WZWLL}:
\begin{eqnarray} \label{fm}
 &&\langle D(p_2)|\bar{q}\gamma_{\mu}(1-\gamma_5)c|J/\psi(\epsilon_{\psi},p_1)\rangle
  \nonumber\\
 &&~~~~~~~=-\epsilon_{\mu\nu\alpha\beta}\epsilon_{\psi}^{\nu}p_1^{\alpha}p_2^{\beta}{2V(q^2)
  \over m_{\psi}+m_{D}}+i(m_{\psi}+m_{D})\left[{\epsilon_{\psi}}_{\mu}-{\epsilon_{\psi} \cdot q
  \over q^2}q_{\mu}\right]A_1(q^2)\nonumber\\
 &&~~~~~~~~~~+i{\epsilon_{\psi} \cdot q \over m_{\psi}+m_{D}}A_2(q^2)\left[(p_1+p_2)_{\mu}
  -{m_{\psi}^2-m_{D}^2\over q^2} q_{\mu}\right]+
  2i m_{\psi}{\epsilon_{\psi} \cdot q \over q^2}q_{\mu}A_0(q^2), \\
 &&\langle D^{*}(\epsilon_{D^{\ast}},p_2)|\bar{q}\gamma_{\mu}(1-\gamma_5)c|J/\psi
 (\epsilon_{\psi},p_1)\rangle \nonumber\\
 &&~~~~~~~=-i\epsilon_{\mu \nu \alpha \beta}\epsilon_{\psi}^{\alpha}\epsilon_{D^{\ast}}^{*\beta}
  \left[(p_1^{\nu}+p_2^{\nu}-{m_{\psi}^2-m_{D^{*}}^2 \over q^2}q^{\nu})
  \tilde{A}_1(q^2)+{m_{\psi}^2-m_{D^{*}}^2 \over q^2}q^{\nu} \tilde{A}_2(q^2)\right]
  \nonumber \\
 &&~~~~~~~~~~+{i \over m_{\psi}^2-m_{D^{*}}^2}\epsilon_{\mu \nu \alpha \beta}
  p_1^{\alpha} p_2^{\beta} [\tilde{A}_3(q^2) \epsilon_{\psi}^{\nu}
  \epsilon_{D^{\ast}}^{*} \cdot q-\tilde{A}_4(q^2) \epsilon_{D^{\ast}}^{*\nu}
  \epsilon_{\psi}
  \cdot q] \nonumber\\
 &&~~~~~~~~~~+(\epsilon_{\psi} \cdot \epsilon_{D^{\ast}}^{*})[-({p_1}_{\mu}+{p_2}_{\mu})
  \tilde{V}_1(q^2)+q_{\mu} \tilde{V}_2(q^2)]\nonumber \\
 &&~~~~~~~~~~+{(\epsilon_{\psi} \cdot q)(\epsilon_{D^{\ast}}^{*} \cdot q) \over
  m_{\psi}^2-m_{D^{*}}^2}\bigg[({p_1}_{\mu}+{p_2}_{\mu}-{m_{\psi}^2-m_{D^{*}}^2
  \over q^2}q_{\mu}) \tilde{V}_3(q^2)\nonumber \\
 &&~~~~~~~~~~+{m_{\psi}^2-m_{D^{*}}^2 \over q^2}q_{\mu}
  \tilde{V}_4(q^2)\bigg]-(\epsilon_{\psi} \cdot q) {\epsilon^{*}_{D^{\ast}}}_{\mu}
  \tilde{V}_5(q^2) + (\epsilon^{*}_{{D^{\ast}} } \cdot q){\epsilon_{{\psi}}}_{ \mu}
  \tilde{V}_{6}(q^2), \label{vector vector}
\end{eqnarray}
where $q=p_1-p_2$ and the convention ${\rm{Tr}}[\gamma_{\mu}
\gamma_{\nu} \gamma_{\rho} \gamma_{\sigma} \gamma_5]=4 i
\epsilon_{\mu \nu \rho \sigma}$ is adopted. For the transition of
$J/\psi$ into a charmed pseudoscalar meson which is induced by the
weak current, there are four independent form factors: $V,~ A_0,~
A_1,~ A_2$; while there are ten form factors for $J/\psi$ transiting
into a charmed vector meson which are parameterized as
$\tilde{A}_{i}(i=1,2,3,4),~\tilde{V}_{j}(j=1,2,3,4,5,6)$.

According to the quark diagrams in Fig. \ref{f1}, the decays are
classified into two categories: color allowed and suppressed
processes.  For the color allowed processes, the decay amplitudes
are proportional to
 \beq
 a_1=C_1+C_2/N_c,
 \eeq
with $N_c$ being the color number of QCD. Because $C_1\sim 1$ and
$C_2\sim \alpha_s$, $a_1$ is estimated to be of order 1. While for
the decays shown in Fig. \ref{f1}(b), the amplitude is proportional
to
 \beq
 a_2=C_2+C_1/N_c.
 \eeq
As $a_2/a_1\sim 1/N_c$, this kind of decays are usually named as
color-suppressed processes.

For the decays of $c\to su\bar d$ which is the Cabibbo-favored
process, the CKM element $V_{cs}V_{ud}$ responsible for these modes
is close to $1$. For the Cabibbo-suppressed transitions of $c\to
du\bar d$ and $c\to su\bar s$, the corresponding CKM parameters
$V_{cd}V_{ud}$ and $V_{cs}V_{us}$ are suppressed by a factor ${\rm
sin}\theta_C\approx 0.22$ with $\theta_C$ being the Cabibbo angle.
The doubly suppressed processes, such as $c\to du\bar s$ which are
suppressed by ${\rm sin}^2\theta_C$, are neglected in our case.
Thus, the  prevailing decay modes are both color allowed and Cabibbo
favored ones. The less dominant modes are the color suppressed but
Cabibbo favored or color allowed but Cabibbo suppressed processes.
These are the processes we will focus on in this study. Note that
there is no annihilation type contributions in our case at all.

Now, we are able to write down the decay amplitudes associating with
the non-leptonic two-body decays of $J/\psi$ explicitly based on the
information we achieved before. In light of the characters of final
states, three different types of processes  $J/\psi\to DP$, $DV$ ,
$D^*P$ and $D^*V$ will be investigated one by one in the following
sections. As for $J/\psi$ decaying into two pseudoscalars where one
is a $D$ meson and the other is a light meson $P$, the decay
amplitude is written as
 \beq
 A(J/\psi\to DP)=\la DP|\mathcal{H}_{eff}|J/\psi\ra=
 {G_{F}\over\sqrt{2}}V^{*}_{cq}V_{uq'}a_i~ 2m_{\psi}
 (\epsilon_\psi\cdot q)f_P A_0(q^2),
 \eeq
where $q$ denotes the momentum of light emitted meson;  $a_i$ is the
effective coefficients with $a_1$ for color allowed process and
$a_2$ for color suppressed process.

The decay amplitude of  $J/\psi\to DV$ decay  is given as
 \beq
 A(J/\psi\to DV)&=&{G_{F}\over\sqrt{2}}V^{*}_{cq}V_{uq'}a_i~ f_Vm_V
  \left\{ -\epsilon_{\mu\nu\alpha\beta}\epsilon^{*\mu}_{V}
  \epsilon_{\psi}^{\nu}p_{\psi}^{\alpha}p_D^{\beta}
  {2V(q^2)\over m_{\psi}+m_{D}}\right. \non \\
  &&\left.
  +i(m_{\psi}+m_{D})(\epsilon_{\psi}\cdot\epsilon_V^*) A_1(q^2)
  +i{(\epsilon_\psi\cdot q)[\epsilon_V^*\cdot (p_1+p_2)]\over m_{\psi}+m_{D}}
  2A_2(q^2)\right \};
 \eeq
 the decay amplitude of  $J/\psi\to D^*P$ decay  can be shown as
\begin{eqnarray}
 A(J/\psi\to D^*P)&=&i{G_{F}\over\sqrt{2}}V^{*}_{cq}V_{uq'}a_i~f_P \bigg\{
 2 i \epsilon_{\mu \nu \alpha \beta}p_1^{\mu} p_2^{\nu}
 \epsilon_{\psi}^{\alpha}\epsilon_{D^{\ast}}^{\ast \beta}
 \tilde{A}_1(q^2) \non \\
 &&+(\epsilon_\psi\cdot\epsilon_{D^*}^{*}) \bigg[
  (m_{\psi}^2-m_{D^*}^2)\tilde{V}_1(q^2)-q^2\tilde{V}_2(q^2) \bigg]
   \non\\
 &&
  +(\epsilon_\psi \cdot q) (\epsilon_{D^*}^*\cdot q) \bigg[
  -\tilde{V}_4(q^2)+\tilde{V}_5(q^2)-\tilde{V}_{6}(q^2) \bigg]
 \bigg  \}.
\end{eqnarray}
Lastly, the expressions for $J/\psi \to D^* V$ can be readily
derived from Eqs. (\ref{dc}, \ref{fm}) as
\begin{eqnarray}
 A(J/\psi\to D^{\ast}V)&=&{G_{F}\over\sqrt{2}}V^{*}_{cq}V_{uq'}a_i~ f_Vm_V
 \bigg \{  -i\epsilon_{\mu \nu \alpha \beta}\epsilon_{\psi}^{\alpha}\epsilon_{D^{\ast}}^{*\beta}
  {\epsilon^{\ast \mu}_{V}}\bigg[(p_1^{\nu}+p_2^{\nu}-{m_{\psi}^2-m_{D^{*}}^2 \over q^2}q^{\nu})
  \tilde{A}_1(q^2) \nonumber \\
  &&~~~~~~~~~~+{m_{\psi}^2-m_{D^{*}}^2 \over q^2}q^{\nu} \tilde{A}_2(q^2)\bigg]
  \nonumber \\
 &&~~~~~~~~~~+{i \over m_{\psi}^2-m_{D^{*}}^2}\epsilon_{\mu \nu \alpha \beta}
  p_1^{\alpha} p_2^{\beta} {\epsilon^{\ast \mu}_{V}} [\tilde{A}_3(q^2) \epsilon_{\psi}^{\nu}
  \epsilon_{D^{\ast}}^{*} \cdot q-\tilde{A}_4(q^2) \epsilon_{D^{\ast}}^{*\nu}
  \epsilon_{\psi}
  \cdot q] \nonumber\\
 &&~~~~~~~~~~-(\epsilon_{\psi} \cdot
 \epsilon_{D^{\ast}}^{*})[{\epsilon^{\ast}_{V}} \cdot (p_1+p_2)
  \tilde{V}_1(q^2)]+{(\epsilon_{\psi} \cdot q)(\epsilon_{D^{\ast}}^{*} \cdot q) \over
  m_{\psi}^2-m_{D^{*}}^2}\bigg[{\epsilon^{\ast}_{V}} \cdot (p_1+p_2) \tilde{V}_3(q^2)\bigg]\nonumber \\
 &&~~~~~~~~~~-(\epsilon_{\psi} \cdot q) {\epsilon^{*}_{D^{\ast}}}
 \cdot {\epsilon^{\ast}_{V}}
  \tilde{V}_5(q^2) + (\epsilon^{*}_{{D^{\ast}} } \cdot
  q){\epsilon_{{\psi}}} \cdot {\epsilon^{\ast}_{V}}
  \tilde{V}_{6}(q^2) \bigg\},
\end{eqnarray}
where ${\epsilon^{\ast}_{V}}$ denotes the polarization vector of
light emitted mesons.

\section{Decay rates for non-leptonic weak decays of $J/\psi$}

\subsection{Input parameters}

The decay rates of the non-leptonic decays $J/\psi\to D+M$ are
written as
\begin{eqnarray}
 \Gamma_{\psi \to DM}={1 \over 3}{1 \over 8\pi}
 \left|A(J/\psi \to DM)\right|^2
 {|{\bf{p}}_D| \over  m_{\psi}^2},
\end{eqnarray}
where ${\bf{p}}_D$ denotes the three-momentum of the final $D$ meson
in the rest frame of $J/\psi$ and the factor ``${1 \over 3}$'' is
due to the spin average of $J/\psi$. In order to calculate the decay
rates, the input parameters including the CKM parameters, effective
Wilson coefficients, decay constants and transition form factors are
necessary. The CKM parameters are taken from ref.\cite{PDG}
 \beq
 V_{ud}=0.974, \qquad V_{us}=0.227, \qquad
 V_{cd}=0.227, \qquad V_{cs}=0.973.
 \eeq
The effective Wilson coefficients are determined as  \cite{Verma 2}
 \beq
 a_1=1.26, \qquad  a_2=-0.51,
 \eeq
which are extracted from the isospin analysis for $D \to K \pi$
decays with the help of the factorization ansatz \cite{wirbel}.

The decay constants for light mesons are taken as
\cite{PDG,Li:2006jv}
 \beq
 f_\pi=0.131 \mbox{ GeV},   \qquad &
 f_K=0.160  \mbox{ GeV},     \qquad & \non\\
 f_\rho=0.209 \pm {0.002}{\mbox{ GeV}} ,   \qquad &
 f_{K^*}=0.217  \pm {0.005}\mbox{ GeV},
 \eeq
where the pseudoscalar  decay constants are determined from the
combined rate for $P \to  l^{\pm} \nu_{l}$ and  $P \to  l^{\pm}
\nu_{l} \gamma$ experimentally and vector meson longitudinal decay
constants are extracted from the data on $\tau^- \to (\rho^-,K^{*-})
\nu_\tau$~\cite{PDG}.

Besides, the values of all the revelent transition form factors
$F_i(q^2)$ are taken from our earlier study \cite{WZWLL} where the
detailed expression are presented. In the literature, the form
factors $A_0$ and $A_1$ have been calculated by various authors
\cite{Verma 1, Verma 2}, which are grouped in the Table~\ref{results
of form factors} together with the numbers obtained in the QCD sum
rules \cite{WZWLL}. We can see that the form factors at zero
momentum transfer predicted in the BSW model \cite{BSW} are
approximately greater than that in the QCD sum rules by a factor 2.

\begin{table}
\caption{The form factors $A_0$ and $A_1$ at $q^2=0$ responsible for
the decays of $J/\psi \to D_{(s)}$ in BSW model \cite{Verma 1} and
QCDSR \cite{WZWLL} approach.} \label{results of form factors}
\begin{ruledtabular}
\begin{tabular}{ccccc}
  Models & $A_0^{\psi \to D}$ & $A_0^{\psi \to D_s}$ & $A_1^{\psi \to D}$ & $A_1^{\psi \to D_s}$
  \\ \hline
  BSW & 0.61 & 0.66 & 0.68 & 0.78  \\ \hline
  QCDSR  & 0.27 & 0.37 & 0.27 & 0.38 \\
\end{tabular}
\end{ruledtabular}
\end{table}

\subsection{Branching ratios of non-leptonic decays}

The numerical results of branching ratios for non-leptonic decays of
$J/\psi \to D_{(s)} P$ are presented in Tables \ref{t1}, where the
numbers obtained in ref.\cite{Verma 1, Verma 2} are also collected
together for a comparison. Here, the results are given for decays
which including the charge conjugate process, for instance,
$BR(J/\psi\to D_s\pi)$ is the branching ratio for decays of
$J/\psi\to D_s^+\pi^-+D_s^-\pi^+$.

\begin{table}
\caption{Branching ratios of non-leptonic decays of $J/\psi\to
D_{(s)}P$ (in units of $10^{-10}$).}\label{t1}
\begin{ruledtabular}
\begin{tabular}{ccc}
                              & other works        & this study    \\ \hline
  $BR(J/\psi\to D_s\pi)$  &    $17.4$ \cite{Verma 1}  &  $2.0^{+0.4}_{-0.2}$ \\
                          &    $10.0$ \cite{Verma 2}  &       \\ \hline
  $BR(J/\psi\to D_sK)$    &    $1.10$ \cite{Verma 1}  &  $0.16^{+0.02}_{-0.02}$ \\ \hline
  $BR(J/\psi\to D\pi)$    &    $1.10$ \cite{Verma 1}  &  $0.080^{+0.02}_{-0.02}$ \\ \hline
  $BR(J/\psi\to DK)$      &    ---  &  $0.36^{+0.10}_{-0.08}$\\
\end{tabular}
\end{ruledtabular}
\end{table}

Table~\ref{t1} shows that the decay rate for color allowed and
Cabibbo favored channel $J/\psi\to D_s \pi$ calculated in this work
is five times smaller than that given in Ref.\cite{Verma 2}. Such
discrepancy may be attributed to two aspects: Firstly, an $SU(4)
f_{ijk}$ rotation matrix  is employed in Ref. \cite{Verma 2} to
relate the $J/\psi\to D_s$ transition to $D \to K^{\ast}$ decay, and
the form factor $A_0(0)$ is estimated as $0.7 \sim 0.8$, which is
almost twice as that computed in the QCD sum rules. Secondly, the
experimental data on total decay width of $J/\psi$ used in Ref.
\cite{Verma 2} is $67.0$ keV, however, this value has been updated
to $93.4 \pm 2.1$ keV \cite{PDG}.

As for the Cabibbo suppressed but color-allowed mode $J/\psi\to D_s
K$, the following relation
\begin{eqnarray}
  R_1 \equiv {  BR(J/\psi\to D_s K) \over BR(J/\psi\to D_s \pi) }
  \approx \left|{V_{us} f_{K}  \over
  V_{ud} f_{\pi} }\right|^2 \approx 0.081
\end{eqnarray}
is achieved in the factorization assumption. Similarly, we can
define a parameter $R_2$ as
\begin{eqnarray}
  R_2 \equiv { BR(J/\psi\to D \pi) \over BR(J/\psi\to D_s \pi) }
  \approx \left|{V_{cd} A_0^{\psi D}(m_{\pi}^2) \over V_{cs} A_0^{\psi D_s}(m_{\pi}^2)}\right|^2 \approx
  0.032,
\end{eqnarray}
which is also in agreement with that listed in Table~ \ref{t1}, as
long as the phase space is properly considered for these two
channels.

Now, we move on to the discussions of color suppressed mode
$J/\psi\to D K$. A ratio of decay rates between it and $J/\psi \to
D_s \pi$ can be estimated as
\begin{eqnarray}
  R_3 \equiv { BR(J/\psi\to D K) \over BR(J/\psi\to D_s \pi) }
  \approx \left|{ a_2 A_0^{\psi D}(m_{K}^2) \over a_1 A_0^{\psi D_s}(m_{\pi}^2)}\right|^2 \approx
  0.18,
\end{eqnarray}
which is consistent with that collected in Table~\ref{t1}. In
addition, we should emphasize that the ratio $R_3$ is quite
sensitive to the effective Wilson coefficient $a_2$, which can
receive considerable corrections \cite{QCDF 1, QCDF 2, PQCD 1, PQCD
2, PQCD 3, SCET 1,SCET 2, h.n. li review} due to uncertainties of
the renormalization scale, higher order effects together with
non-factorizable contributions, where we also refer to \cite{Beneke}
for a  recent comment.

\begin{table}
\caption{Branching ratios of non-leptonic decays of $J/\psi\to
D_{(s)}^* P$ and $D_{(s)}V$  (in units of $10^{-10}$).}\label{t2}
\begin{ruledtabular}
\begin{tabular}{ccc}
                            & other works        & this study    \\ \hline
  $BR(J/\psi\to D_s\rho)$   &   $72.6$ \cite{Verma 1}   &  $12.6^{+3.0}_{-1.2}$ \\ \hline
  $BR(J/\psi\to D_sK^*)$    &   $4.24$ \cite{Verma 1}   &   $0.82^{+0.22}_{-0.10}$\\ \hline
  $BR(J/\psi\to D\rho)$     &   $4.40$ \cite{Verma 1}  &   $0.42^{+0.18}_{-0.08}$\\ \hline
  $BR(J/\psi\to DK^*)$      &   ---    &   $1.54^{+0.68}_{-0.38}$\\ \hline \hline
  $BR(J/\psi\to D_s^*\pi)$  &   ---    &   $15.0^{+1.2}_{-0.4}$\\ \hline
  $BR(J/\psi\to D_s^*K)$    &   ---    &   $1.1^{+0.08}_{-0.04}$\\ \hline
  $BR(J/\psi\to D^*\pi)$    &   ---    &   $0.60^{+0.04}_{-0.04}$\\ \hline
  $BR(J/\psi\to D^*K)$      &   ---    &   $2.6^{+0.2}_{-0.2}$\\
\end{tabular}
\end{ruledtabular}
\end{table}

Table \ref{t2} collects the numerical results for $J/\psi\to
D_{(s)}^* P$ and $D_{(s)}V$ decays. As one can see, the branching
ratio of $J/\psi\to D_s \rho$ computed in \cite{Verma 1} is $5.8$
times larger than that evaluated in this work. It is shown in
\cite{Verma 1}, the dominant contributions for decay width of $B \to
D^{(\ast)} P$ are from the form factor $A_1(q^2)$ corresponding to
the S partial wave in the final states. As listed in Table.
\ref{results of form factors}, the number of form factor $A_1$
derived in the BSW model is $2.1$ times greater than that in terms
of the QCD sum rules, which can indeed result in an enormous
discrepancy for the branching fraction of $J/\psi\to D_s$ obtained
in two different approaches. Moreover, the ratio
\begin{eqnarray}
  R_4 \equiv {  BR(J/\psi\to D_s \rho) \over BR(J/\psi\to D_s \pi) },
\end{eqnarray}
is usually introduced from a viewpoint of experiment, whose value is
estimated as $6.3$ and $4.2$ respectively in the framework of QCDSR
and BSW model. Therefore, the decay of $J/\psi\to D_s \rho$ is more
detectable than the corresponding pseudoscalar channel $J/\psi\to
D_s \pi$ in experiment. Moreover, it can be seen that the ratio of
decay rates is not sensitive to the absolute magnitude of the
transition form factors on account of the large cancelations of the
non-perturbative effects. The relative magnitude of decay widths for
the Cabibbo-suppressed as well as color suppressed processes to the
mode of $J/\psi\to D_s \rho$ can  be readily derived by following
the discussions on $J/\psi \to D_{(s)} P$ and will not be repeated
again.

\begin{table}
\caption{Branching ratios of non-leptonic decays of $J/\psi\to
D^*_{(s)} V$ (in units of $10^{-10}$).}\label{t3}
\begin{ruledtabular}
\begin{tabular}{cc}
             Channels            & this study    \\ \hline
  $BR(J/\psi\to D_s^*\rho)$     &   $52.6^{+7.2}_{-6.2}$\\ \hline
  $BR(J/\psi\to D_s^*K^*)$         &   $2.6^{+0.4}_{-0.4}$\\ \hline
  $BR(J/\psi\to D^*\rho)$        &   $2.8^{+0.6}_{-0.4}$\\ \hline
  $BR(J/\psi\to D^*K^*)$           &   $9.6^{+3.2}_{-2.2}$\\
\end{tabular}
\end{ruledtabular}
\end{table}

Furthermore, we group the decay rates for dominant channels of
$J/\psi\to D^*_{(s)} V$ in Table~ \ref{t3}, from which we can
observe that $BR(J/\psi\to D_s^*\rho)$ is as large as $5.3 \times
10^{-9}$ and stands as the most promising mode to be measured at
BESIII. Such finding presents a striking contrast to the argument
given by the authors of Ref. \cite{Sanchis-Lonzano} where the
authors claimed that a specific non-leptonic decay channel like
$J/\psi \to D_{s}^{(*)} M$ ($M=\pi, \rho ...$) is hardly to be
detected owing to the tiny branching factions  for these processes.
In addition, it is also helpful to define the following two ratios
\begin{eqnarray}
R_5 \equiv {{\rm{BR}}(J/\psi \to D^{*}_{s} \pi) \over
{\rm{BR}}(J/\psi \to D_{s} \pi)}, \qquad R_6 \equiv
{{\rm{BR}}(J/\psi \to D^{*}_{s} \rho) \over {\rm{BR}}(J/\psi \to
D_{s} \rho)},
\end{eqnarray}
which   characterize  the relative size of branching fractions to
distinguish the final states with vector and pseudoscalar ones
respectively in the non-leptonic two-body weak decays of $J/\psi$.
The numbers of $R_5$ and $R_6$ are evaluated as $7.5$ and $4.2$ in
the QCD sum rules, while they are determined as $3.5$ and $1.4$
respectively with the ISGW model in the framework of heavy quark
spin symmetry \cite{Sanchis-Lonzano}. Such discrepancies can be
attributed to the different values of form factors employed in the
numerical calculations.

Moreover, we  also mention that
\begin{eqnarray}
R_7 \equiv {{\rm{BR}}(J/\psi \to D^{*}_{s} K^{\ast}) \over
{\rm{BR}}(J/\psi \to D^{\ast} \rho)} \cdot {{\rm{BR}}(J/\psi \to D
\rho) \over {\rm{BR}}(J/\psi \to D_s K^{\ast})}
\end{eqnarray}
should be equal to $1$ in the heavy quark limit. However, this ratio
is estimated as $0.48$  in the QCD sum rules owing to a serious
suppression factor from the phase space for the decay of $J/\psi \to
D^{*}_{s} K^{\ast}$ for the limited charm quark mass.

Combining the Table~\ref{t1}, \ref{t2} and \ref{t3}, we find that
the branching ratio for inclusive weak decay of $J/\psi$ can be as
large as $1.3 \times 10^{-8}$, which is also in remarkable agreement
with the naive estimation
\begin{eqnarray}
BR(J/\psi \to X_c +...) \approx {2 \Gamma_{D^{\pm}} \over
\Gamma_{J/\psi}} \approx 1.4 \times 10^{-8}.
\end{eqnarray}

\section{Discussions and conclusions}
\label{Discussions and conclusions}

Since $J/\psi$ mainly decays via strong and electromagnetic
interactions, its weak decays usually take small fractions which
cannot be measured by available experimental apparatus. On other
aspect, however, because $J/\psi$ contains two heavy constituents,
its weak decay may possess a unique character. Indeed weak decays of
$J/\psi$ may offer an ideal platform to examine the  mechanism which
governs the hadronization process, without possible contamination
from the light spectator as well as one may determine its
fundamental parameters such as the CKM matrix which can be a
complementary test to the values obtained in $D$ decays. It is lucky
for   high energy physicists  that a tremendous database on $J/\psi$
will be available in the forthcoming BESIII and the measurements on
the weak decays of $J/\psi$ may become possible.

As is well known, the essential challenge in the theoretical
calculations on the rates of weak decays of $J/\psi$ is to
disentangle the underlying weak-interaction transitions from the
notorious effects owing to strong interactions reasonably
\cite{neubert}. In our previous paper \cite{WZWLL}, the transition
form factors in the semi-leptonic weak decays of $J/\psi$ have been
investigated to the leading order of $\alpha_s$ based on QCD sum
rules, where the non-perturbative QCD dynamics is characterized by a
few universal parameters. The branching ratios for dominant
exclusive processes are evaluated and their order of magnitude is
typically at $10^{-10}$. Obviously based on the factorization
assumption, the form factors obtained for the semi-leptonic decays
can be applied to study the non-leptonic decays.

This paper can be viewed as a continuation of  our earlier work
\cite{WZWLL}. We present a comprehensive study of non-leptonic
decays of $J/\psi \to  D_{(s)}+M$ based on the factorization
assumption and apply the transition form factors calculated in the
QCD sum rules. It is observed that the sum of the branching
fractions for the dominant non-leptonic decays of $J/\psi \to
D^{-}_{s} \pi$, $ D^{-}_{s} \rho$, $D^{*-}_{s} \pi$, and $D^{*-}_{s}
\rho$ as well as their charge conjugate  channels can reach as large
as $0.82 \times 10^{-8}$, a special decay mode $J/\psi \to D^{*}_{s}
\rho$ can even arrive at $5.3 \times 10 ^{-9}$, which is hopefully
to be marginally detected in the $e^{+} e^{-}$ colliders in view of
the large database of the BESIII. Our results are in agreement with
the finding in Ref.\cite{Sanchis-Lonzano} that $J/\psi$ decays to
vector charmed meson $D_s^{*-}$ more favorably than to the
pseudoscalar one, however, the ratios of these two channels
calculated in this work is twice or three times larger than that
given by Ref.\cite{Sanchis-Lonzano}, where heavy quark spin symmetry
and the non-recoil approximation were adopted and the ISGW model was
employed to compute the single form factor $\eta_{12}$.

As there is a light to see a possibility of measuring weak decays
of $J/\psi$ which have obvious advantages for getting insight to
the physics picture, we strongly urge our experimental colleagues
to search for vector charmed mesons productions in $J/\psi$ decays
at BESIII \cite{BES-semi,BES-non}.

Moreover, from the theoretical side, it should be emphasized that
Coulomb-type corrections for the  heavy quarkonium system
\cite{QCDSR 1,kiselev, Coulomb corrections, Kiselev prd} are not
included in the computations of form factors in the QCD sum rues,
which could induce additional uncertainties to the evaluation of the
branching fractions for non-leptonic two body decays of $J/\psi$.
However, one can still trust the order of magnitude gained in this
work, since while calculating the form factors which need to deal
with the three-point correlations, most uncertainties originating
from Coulomb-like corrections are canceled by that in the two-point
correlations for evaluating the decay constant of $J/\psi$. Apart
from weak decays of $J/\psi$ presented in this paper, weak decays of
$\Upsilon$ are suggested to be explored seriously in a complementary
fashion from both the theoretical and experimental point of view.

\section*{Acknowledgements}

This work is partly supported by National Science Foundation of
China under Grant Nos. 10735080, 10625525, 10705015 and 10745002.

\end{document}